\documentclass{iau}
\usepackage{graphicx}
\usepackage{natbib}

\title[Features of the CON Galaxy NGC4418] 
{Large-Scale Features of the CON \\ Galaxy NGC4418 with MUSE}

\author[Clare F. Wethers]   
{Clare F. Wethers$^1$
 Susanne Aalto$^1$
 G. C. Privon $^{2,3,4}$
 F. Stanley$^5$
 J. Gallagher$^6$
 M. Gorski$^1$
 S. König$^1$
 K. Onishi$^1$
 C. Yang$^1$}

\affiliation{$^1$Department of Space, Earth and Environment, Chalmers University of Technology, \\ 
            Onsala Space Observatory, 439 92 Onsala, Sweden   \\[\affilskip]
            
            $^2$National Radio Astronomy Observatory, Charlottesville, VA 22903, USA  \\[\affilskip]
            
            $^3$Department of Astronomy, University of Florida, Gainesville, FL 32611, USA  \\[\affilskip]
            
            $^4$Department of Astronomy, University of Virginia, Charlottesville, VA 22904, USA  \\[\affilskip]
            
            $^5$IRAM, Domaine Universitaire
            38406 Saint Martin d’Hères, France \\[\affilskip]
            
            $^6$Wisconsin IceCube Particle Astrophysics Center, Madison, WI 53703, USA}

\pubyear{2023}
\volume{xxx}  
\pagerange{119--126}
\jname{Title of your IAU Symposium}
\editors{A.C. Editor, B.D. Editor \& C.E. Editor, eds.}
\begin{document}

\maketitle

\begin{abstract}

Compact obscured nuclei (CONs) are relatively common in the centers of local (U)LIRGs, yet their nature remains unknown. Both AGN activity and extreme nuclear starbursts have been suggested as plausible nuclear power sources. The prevalence of outflows in these systems suggest that CONs represent a key phase in the nuclear feedback cycle, in which material is ejected from the central regions of the galaxy. Here, we present results from MUSE for the confirmed local CON galaxy NGC4418. For the first time we spatially map the spectral features and kinematics of the galaxy in the optical, revealing several previously unknown structures. In particular, we discover a bilateral outflow along the minor axis, an outflowing bubble, several knot structures and a receding outflow partially obscured by the galactic disk. Based on the properties of these features, we conclude that the CON in NGC4418 is most likely powered by an AGN.


\end{abstract}

\firstsection 
\section{Introduction}

Compact obscured nuclei (CONs) exist in 20-40\% of nearby (ultra-) luminous infrared galaxies (ULIRGs) and are primarily characterized by their extreme nuclear column densities, n$_{H2}$ $>$10$^{25}$cm$^{-2}$. This extreme attenuation renders them almost invisible at mid-IR, optical and X-ray wavelengths, which, along with their compact size ($\lesssim$100 pc), makes CONs notoriously difficult to find. To date, the only method by which these objects can be identified is via a rare transition of vibrationally excited HCN, $\nu_{2}$ = 1f (HCN-VIB), which traces hot dust at high column densities \citep{aalto15b}. Although only a handful of CONs have been confirmed, all known CONs show signatures of a past gas-rich merger and evidence for molecular outflows in HCN and/or CO. The prevalence of outflows in CONs is consistent with a heavily obscured phase in which remnant material is being ejected from the center of the galaxy. CONs are therefore critical in understanding the role of nuclear feedback in galaxy evolution, yet their nature and what powers their outflows remains unknown. 

On the one hand, CONs may denote an extreme nuclear nascent starburst deeply embedded in dust. While this scenario would explain the unusual FIR and radio properties of CONs \citep[e.g.][]{yun01,roussel03}, these starbursts are rare, and several studies have argued that dust could not obscure an extended starburst to the extent required to explain the deep absorption features observed in CONs \citep[e.g.][]{roche86, spoon01}. This has led to the idea that CONs may instead denote a class of hidden active galactic nuclei (AGN). In such a scenario, the heavy nuclear attenuation likely arise following a rapid accretion event such as a merger or galaxy-galaxy interaction \citep[e.g.][]{kocevski15, ricci17, boettcher20}. The likelihood of such interaction events have been shown to increase with (U)LIRG luminosity \citep{sanders96}, meaning if CONs are associated with rapid gas inflow they too should preferentially exist in more luminous systems. One of the key pieces of evidence to support the AGN scenario is therefore the increased CON fraction in ULIRGs (10$^{11}<$L$_{\rm{IR}}>$10$^{12}$ L$\odot$; $\sim$40 per cent) compared to LIRGs (L$_{\rm{IR}}>$10$^{11}$ L$\odot$; $\sim$20 per cent). 

Directly distinguishing between the starburst and hidden AGN models of CONs is challenging as many of the traditional AGN signatures in the spectra are completely erased by the extreme nuclear attenuation. We therefore rely on indirect tracers of AGN and/or starburst activity to discover what powers CONs and in turn what role they play in the coevolution of galaxies and their central super-massive black holes (SMBHs). 

\begin{figure}[b]
\begin{center}
 \includegraphics[trim= 50 70 30 50,clip,width=\textwidth]{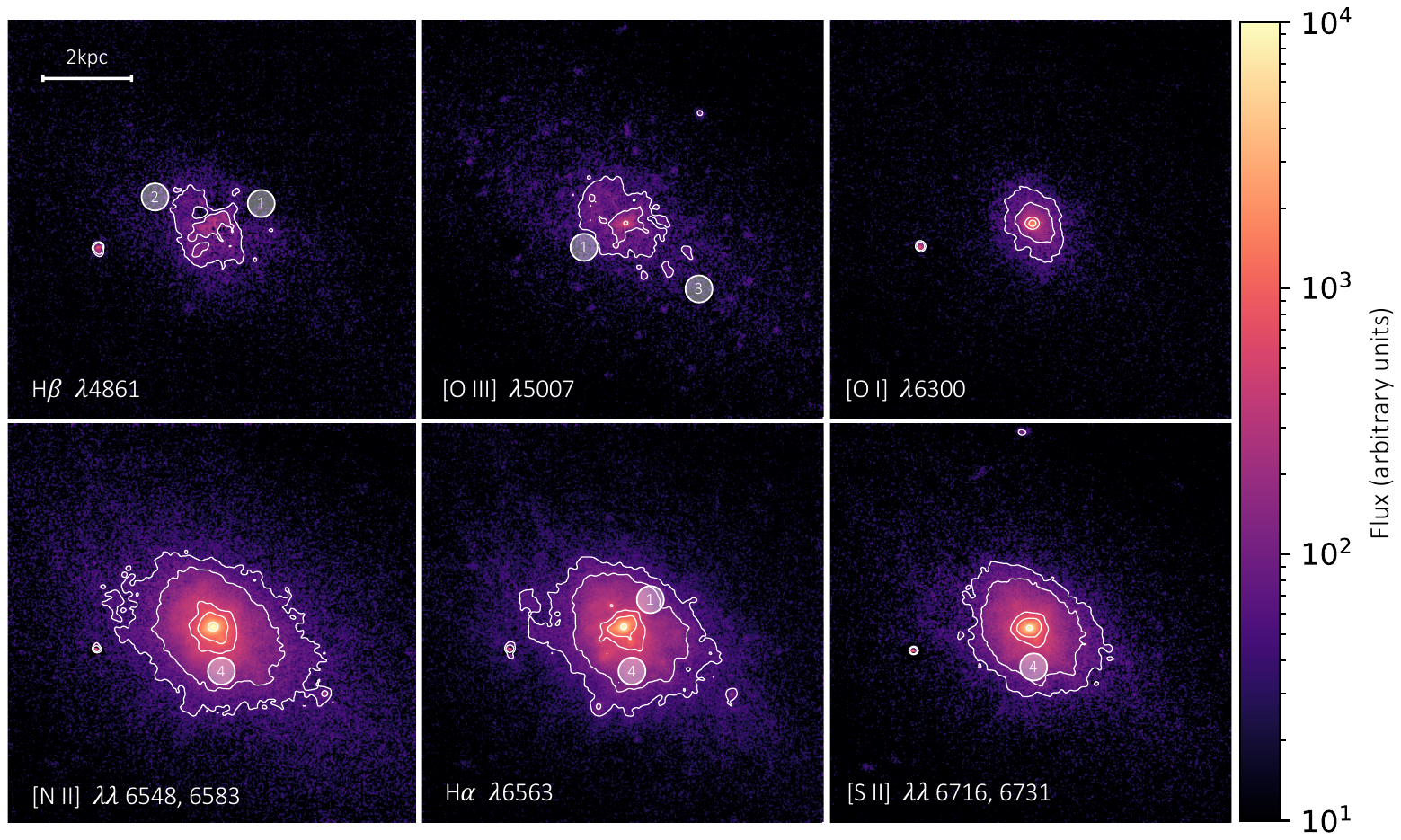} 
 \caption{Integrated line maps of the key spectral features in NGC4418 with contour levels overlaid. Several prominent features of the maps are labelled (1: minor axis elongation, 2: bubble structure, 3: knots, 4: southern outflow) and are discussed in section 2.}
   \label{fig1}
\end{center}
\end{figure}

\begin{figure}[b]
\begin{center}
 \includegraphics[trim= 210 220 200 230,clip,width=\textwidth]{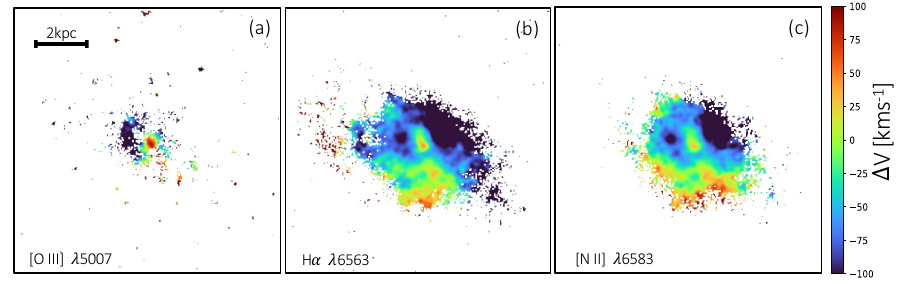} 
 \caption{Moment-1 velocity maps for [O III], H$\alpha$ and [N II], showing the kinematic structure of NGC4418. A 3$\sigma$ signal-to-noise cut and 1$\sigma$ gaussian smoothing have been applied to improve the visual output.}
   \label{fig2}
\end{center}
\end{figure}

\begin{figure}[b]
\begin{center}
 \includegraphics[trim= 0 70 50 90,clip,width=\textwidth]{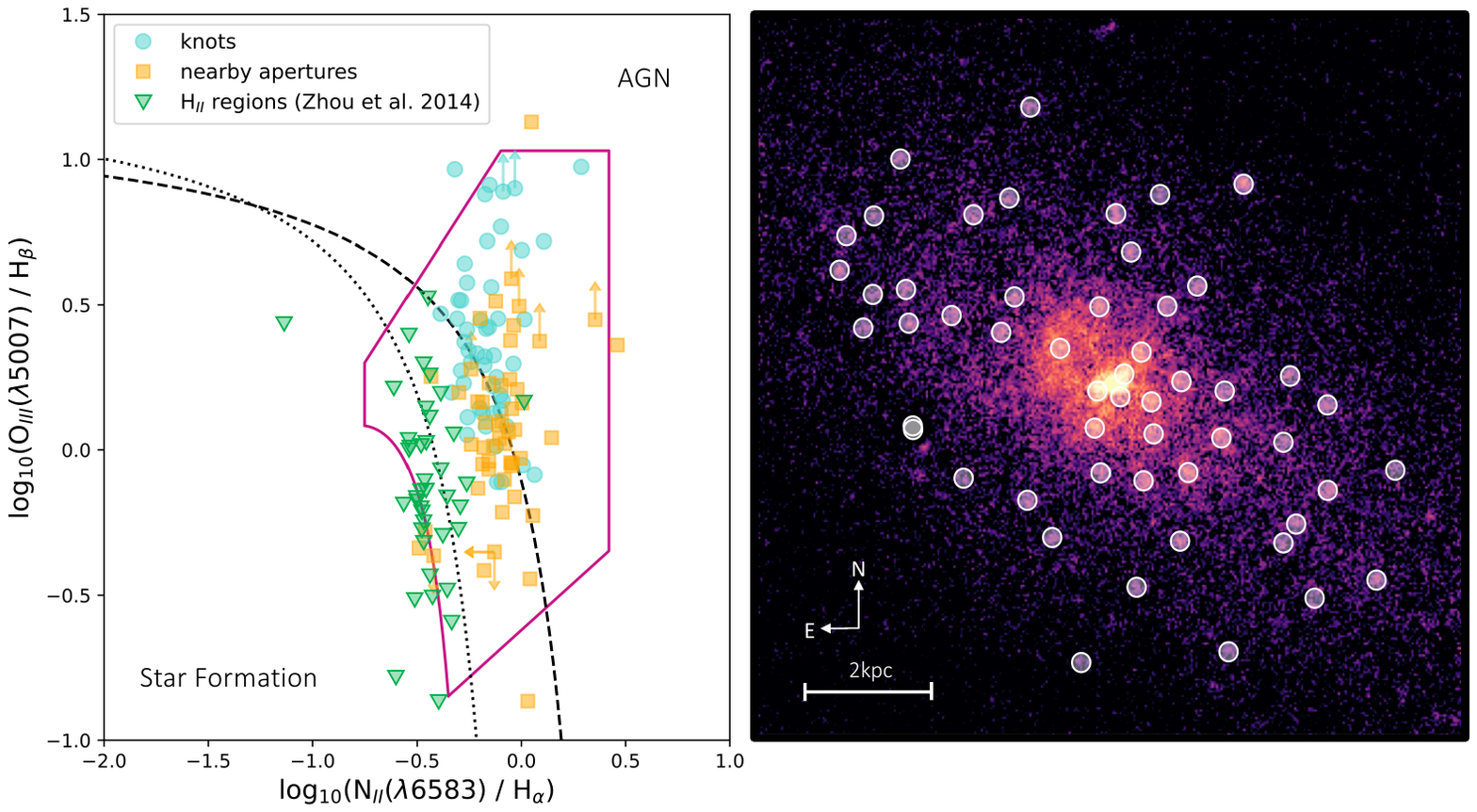} 
 \caption{\textit{Left:} BPT diagram separating ionisation from star formation and AGN. Regions are defined by the models of \cite{kauffmann03} (dotted) and \cite{kewley06} (dashed). Line ratios are measured from integrated aperture spectra (0.4 arcsec radius) centred on the knots (blue circles) and nearby galaxy regions (yellow squares), and compared to a sample of H$_{\rm{II}}$ regions \citep{zhou14}. The pink region shows ratios consistent with shock heating \citep{alatalo16}. \textit{Right:} map of the [O III] line emission with the location of the knots overlaid. }
   \label{fig3}
\end{center}
\end{figure}

\section{Overview}

We present the first results from Wethers et al. \textit{in prep.}, based on new targeted observations from the Multi-Unit Spectroscopic Explorer (MUSE). The work focuses on the local CON galaxy NGC4418 ($z$=0.00727). Our key results are as follows;

{\underline{\it 1. An elongation along the minor axis}:} This structure is present in all of the emission features (Fig.1), but is most noticeably traced by the H$\beta$ and [O III] emission. The angle of the observed elongation is consistent with the minor axis outflow previously observed in NGC4418 by \cite{ohyama19}. This outflow was found to extend to northwest of the galaxy nucleus and was catagorized as a dusty superwind due to its shock-heated emission lines and enhanced stellar NaD absorption. Unlike the outflow detected by \cite{ohyama19} however, we find the elongation to extend in both directions from the galaxy nucleus. Kinematic measurements (Fig.2) reveal this structure to be strongly blueshifted in both directions with a line of sight velocity, $v_{\rm{los}}$ $\gtrsim$ 100 km$^{-1}$.

{\underline{\it 2. A shell/ bubble structure}:} To the north of the nucleus, we detect a bubble-like structure in the H$\beta$ emission (Fig.1). The structure itself appears similar to that observed in the so-called \textit{teacup galaxy}, J1430+1339, which \cite{gagne14} propose to be a fading AGN candidate. 

{\underline{\it 3. Knots throughout the galaxy}:} One of the most striking findings of this work is the presence of tens of knot structures throughout the galaxy, traced by the [O III] emission. We conclude these structures to be so-called \textit{AGN-echos}: regions of the galaxy that have been photoionized by an AGN. This conclusion is drawn primarily from three key pieces of evidence. i) Comparisons to HST V- (F336SW) and U-band (F555W) imaging (Gallagher et al. \textit{in prep.}) reveal no optical counterpart to the knots observed with MUSE. This counterpart would be expected if the knots were tracing regions of the galaxy ionised by young stars e.g. stellar clusters or H$_{II}$ regions. ii) Kinematic measurements (Fig.2) suggest at least some fraction of the knots are kinematically independent from the disk rotation of the galaxy. This rules out the possibility that the knots arise from planetary nebulae and further eliminates the stellar cluster scenario, as both would result in a similar rotational motion of the knots with the stellar disk of the galaxy. iii) Placing the knots on a Baldwin, Phillips and Terlevich (BPT) diagram (Fig.3) reveals the knots to have higher [O III]/H$\beta$ ratios than both nearby regions of the galaxy and H$_{\rm{II}}$ regions \citep{zhou14}. All the knots are found to be consistent with ionization from an AGN and/ or shock heating \citep{alatalo16}. 

{\underline{\it 4. A southern outflow}:} The emission of NGC4418 shows extended emission to the south of the nucleus (Fig.1). Kiematic measurements (Fig.2) show this structure to be redshifted with $v_{\rm{los}}\sim$50 kms$^{-1}$. We therefore postulate that this structure is likely an outflow orientated behind the galaxy disk.

\section{Implications}

The detection of multiple outflow structures in NGC4418 is consistent with the galaxy existing in an early phase of a feedback cycle in which dust is being expelled from the nuclear region via outflows. Based on the velocity and structure of the observed outflows, we suggest that it is unlikely that the CON in NGC4418 is powered solely by starburst activity. Furthermore, the knots detected in the [O III] emission are consistent with having been ionized by an AGN. We therefore conclude that the CON in NGC4418 is likely undergoing an AGN phase. Further future IFU studies are required to determine whether this classification holds for the wider CON population, or whether NGC4418 is unique in this regard. If CONs are indeed powered by AGN, this will double the number of AGN predicted to exist in the local universe \citep{maiolino03} and thus have severe implications for our understanding of AGN-galaxy coevolution. 
\vspace{2pt}

\noindent
\textit{Acknowledgements:} CW acknowledges support from the ERC Advanced grant 789410.

\bibliographystyle{aa}
\bibliography{ref}
\end{document}